\documentstyle[multicol,aps,prb,epsf]{revtex} 

\begin{document}
\title{Size effects on generation-recombination noise}
\author{G. Gomila}
\address{Centre de Recerca en Bioelectr\`{o}nica i 
Nanobioci\`{e}ncia and Departament d'Electr\`{o}nica-Universitat de
Barcelona,
Laboratori de Nanobioingenyieria, Parc Cient\'{\i}fic de Barcelona,
C/ Josep Samitier 1-5, E-08028 Barcelona, Spain}
\author{L. Reggiani}
\address{INFM-National Nanotechnology Laboratory and Dipartimento di Ingegneria
dell'Innovazione,
Universit\`{a} di Lecce, Via Arnesano s/n, I-73100 Lecce, Italy}
\maketitle

\begin{abstract}
We carry out an analytical theory of generation-recombination noise for a
two level resistor model which goes beyond those presently available by
including the effects of both space charge fluctuations and diffusion
current. Finite size effects are found responsible for the saturation of the
low frequency current spectral density at high enough applied voltages. The
saturation behaviour is controlled essentially by the correlations coming
from the long range Coulomb interaction. It is suggested that the saturation
of the current fluctuations for high voltage bias constitutes a general
feature of generation-recombination noise.
\end{abstract}



\smallskip 
\begin{multicols}{2}
Generation recombination (GR) noise is due to fluctuations in the number of
free carriers inside a device and is associated with random transitions of
charge carriers between states in different energy bands. Accordingly, it
represents an important noise source in semiconductor devices where carrier
concentration can vary over many orders of magnitude. Typical examples of
transitions are those between the conduction band and localized levels in
the energy gap, conduction and valence bands, etc. GR noise can be detected
as an excess noise when measuring current or voltage fluctuations.\cite
{Vliet65,Vliet75,Ziel86,Bonani01}

The calculation of the number fluctuations, and hence of the current and
voltage fluctuations, due to GR processes constitutes in general a rather
complicated problem. Usually, one has to resort to simplifying assumptions
as for instance the local space charge neutrality approximation and/or the
neglect of the diffusion current.\cite{Vliet65,Vliet75}The physical
implications of these approximations remain still an unsolved problem, since
an exact theory including simultaneously both effects is lacking in the
current literature.

The aim of the present paper is to address this issue by providing an exact
analytical solution of the low frequency GR noise properties of a two-level
resistor model within the framework of the drift-diffusion transport theory.
The main result of the exact solution is the evidence that size effects on
GR noise can lead to the saturation of the associated current spectral
density at high enough applied bias. Remarkably, present results confirm
unexpected findings of recent numerical studies by Bonani and Ghione\cite
{Bonani99} concerning uniformly $n-$doped silicon samples with band-to-band
and trap assisted generation-recombination processes, where the low
frequency spectral density of the voltage fluctuations was shown to become
independent of the applied bias for moderately strong applied bias.

We consider a two terminal structure consisting of a uniformly $n-$doped
semiconductor sample of length $L$ terminated by (metallic) ohmic contacts
with a single trap (donor) level controlling the electron concentration. We
assume that the transversal dimensions of the sample are much greater than
the longitudinal dimension, thus allowing for a one dimensional
electrostatic treatment. Without loss of generality, only GR noise sources
are considered. The physical model necessary to describe the noise
properties of this system consists of the continuity equations for the free
electrons and the ionized traps, the current equation for the electrons, the
Poisson equation and appropriate boundary conditions. The continuity
equations for free carriers and ionized traps densities read\cite
{Vliet65,Vliet75,Bonani01} 
\begin{eqnarray}
&&\frac{\partial n(x,t)}{\partial t}=\frac{1}{q}\frac{\partial J_{n}(x,t)}{%
\partial x}-r_{n}(x,t)+g_{n}(x,t)+\gamma (x,t)\text{,}  \label{n_cont} \\
&&\frac{\partial N_{t}^{+}(x,t)}{\partial t}=-r_{n}(x,t)+g_{n}(x,t)+\gamma
(x,t)\text{,}  \label{Nt_cont}
\end{eqnarray}
where $n(x,t)$ and $N_{t}^{+}(x,t)$ refers to the free electron and ionized
trap densities at point $x$ and time $t$, respectively, $q$ is the positive
electron charge, $J_{n}(x,t)$ the conduction electron current, $r_{n}(x,t)$
the recombination rate and $g_{n}(x,t)$ the generation rate. Furthermore, $%
\gamma (x,t)$ is the Langevin noise source related to GR processes which has
zero mean and low frequency spectral density\cite{Vliet65,Vliet75,Bonani01} 
\begin{equation}
2\int\limits_{-\infty }^{+\infty }dt\overline{\gamma (x,t)\gamma
(x,t^{\prime })}=\frac{2}{A}\left[ \overline{g}_{n}(x)+\overline{r}%
_{n}(x)\right] \delta (x-x^{\prime })\text{,}  \label{gg}
\end{equation}
with $A$ being the sample cross-sectional area. The bar denotes average with
respect to fluctuations. In the drift-diffusion approach, the conduction
current writes: 
\begin{equation}
J_{n}(x,t)=q\mu n(x,t)E(x,t)+qD_{n}\frac{\partial n(x,t)}{\partial x}\text{,}
\label{Jn}
\end{equation}
where $\mu $ is the electron mobility, $E(x,t)$ the electric field and $%
D_{n} $ the diffusion coefficient satisfying Einstein relation, $D_{n}=\mu
_{n}k_{B}T/q$, with $k_{B}$ being Boltzmann's constant, $T$ the temperature
and where non-degenerate statistics is assumed. To perform a self-consistent
calculation, we take into account Poisson equation 
\begin{equation}
\frac{\partial E(x,t)}{\partial x}=\frac{q}{\epsilon }\left[
N_{t}^{+}(x,t)-n(x,t)\right] \text{,}  \label{Poisson}
\end{equation}
where $\epsilon $ is the semiconductor dielectric permittivity. The total
electric current is thus given by 
\begin{equation}
I(t)=\frac{A}{L}\int_{0}^{L}dx\left[ J_{n}(x,t)+\epsilon \frac{\partial
E(x,t)}{\partial t}\right] \text{.}  \label{I_tot}
\end{equation}
Finally, as appropriate boundary conditions we take ohmic contacts made of
ideal metal-semiconductor interfaces in the diffusion approximation, i.e. we
assume that the diffusion velocity is much smaller than the contact
recombination velocity, thus allowing to neglect any contribution from
thermionic emission processes.\cite{Gomila98} Accordingly, we take 
\begin{equation}
\overline{n}(0)=\overline{n}(L)\equiv n_{c}\text{; \quad }\delta
n(0,t)=\delta n(0,t)=0\text{,}  \label{bc}
\end{equation}
where the carrier density at the contact is 
\begin{equation}
n_{c}=N_{C}e^{-\frac{q\phi _{bn}}{k_{B}T}}\text{,}  \label{n_c}
\end{equation}
with $N_{C}$ being the effective density of states in the conduction band
and $\phi _{bn}$ the contact barrier height. Here, $\delta n(x,t)$ refers to
the free electron density fluctuation.

The model presented above admits an homogeneous stationary solution in the
form 
\begin{equation}
\overline{n}(x)=\overline{N}_{t}^{+}(x)=n_{c}\text{; \quad }\overline{E}(x)=%
\frac{\overline{I}}{qA\mu \overline{n}}\text{ ;}  \label{n_hom}
\end{equation}
Note that being the density of traps $N_{t}$ and the contact density $n_{c}$
two independent variables, one can always find for them values which satisfy
simultaneously Eq. (\ref{n_hom}) and $\overline{g}_{n}(\overline{n},%
\overline{N}_{t}^{+})=\overline{r}_{n}(\overline{n},\overline{N}_{t}^{+})$.
Under these homogeneous conditions the system behaves as a resistor with
current-voltage characteristics satisfying Ohm's law, $\overline{I}=V/R$,
with resistance $R=L/(qA\mu \overline{n})$.

To calculate the low frequency number fluctuations 
\begin{equation}
S_{N}(0)=2\int_{-\infty}^{+\infty}\overline{\delta 
N(t)\delta N(t^{\prime })}dt^{\prime }%
\text{,}  \label{SN_def}
\end{equation}
we use that 
\begin{equation}
\delta N(t)=A\int_{0}^{L}dx\delta n(x,t)\text{,}  \label{dN_def}
\end{equation}
where $\delta n(x,t)$ for low frequencies satisfies 
\begin{equation}
\frac{\partial ^{2}\delta n}{\partial x^{2}}+\frac{1}{L_{E}}\frac{\partial
\delta n}{\partial x}-\frac{1}{\widetilde{L}_{D}^{2}}\delta n=-\frac{\tau }{%
\widetilde{L}_{D}^{2}}\gamma \text{.}  \label{dnx_eq}
\end{equation}
Equation (\ref{dnx_eq}) can be derived by combining Eqs. (\ref{n_cont}), (%
\ref{Nt_cont}), (\ref{Jn}) and (\ref{Poisson}) after linearization around
the homogeneous stationary state and by neglecting the time derivatives due
to the low frequency assumption. The parameters in Eq.(\ref{dnx_eq}) are
defined as 
\begin{equation}
\frac{L_{E}}{L}=\frac{k_{B}T}{qV}\text{; }\widetilde{L}_{D}=\sqrt{\frac{%
k_{B}T\epsilon }{q^{2}\overline{n}(1+\frac{\tau _{t}}{\tau _{n}})}}\text{; }%
\frac{1}{\tau }=\frac{1}{\tau _{t}}+\frac{1}{\tau _{n}}\text{.}
\label{param}
\end{equation}
Here, $\widetilde{L}_{D}$ is a renormalized Debye screening length and 
\begin{eqnarray}
\frac{1}{\tau _{n}} &=&\frac{\partial r_{n}(n,N_{t}^{+})}{\partial n}-\frac{%
\partial g_{n}(n,N_{t}^{+})}{\partial n}\text{;}\quad  \label{taun} \\
\frac{1}{\tau _{t}} &=&\frac{\partial r_{n}(n,N_{t}^{+})}{\partial N_{t}^{+}}%
-\frac{\partial g_{n}(n,N_{t}^{+})}{\partial N_{t}^{+}}\text{.}  \label{taut}
\end{eqnarray}
From the solution of the second order differential equation (\ref{dnx_eq})
we calculate $\delta N(t)$ through Eq. (\ref{dN_def}), and after some
lenghtly but straightforward algebra we arrive at the following expression
for $S_{N}(0)$, 
\begin{equation}
S_{N}(0)=S_{N}^{\infty }(0)-S_{N}^{ex}(0)\text{,}  \label{SN_sol}
\end{equation}
where 
\begin{equation}
S_{N}^{\infty }(0)=4AL\overline{g}_{n}\tau ^{2}=4\left\langle \Delta
N^{2}\right\rangle \tau \text{,}  \label{SN_inf}
\end{equation}
and 
\begin{eqnarray}
S_{N}^{ex}(0) &=&4\left\langle \Delta N^{2}\right\rangle \tau \frac{\left(
e^{\lambda _{1}L}-1\right) \left( e^{\lambda _{2}L}-1\right) \left( \lambda
_{1}-\lambda _{2}\right) }{2L\lambda _{1}\lambda _{2}\left( \lambda
_{2}+\lambda _{1}\right) \left( e^{\lambda _{2}L}-e^{\lambda _{1}L}\right)
^{2}}\times  \nonumber \\
&&\left[ \lambda _{1}\left( 1-e^{(\lambda _{1}+\lambda _{2})L}-3e^{\lambda
_{2}L}+3e^{\lambda _{1}L}\right) \right.  \label{SN_ex} \\
&&\left. -\lambda _{2}\left( 1-e^{(\lambda _{1}+\lambda _{2})L}-3e^{\lambda
_{1}L}+3e^{\lambda _{2}L}\right) \right] \text{.}  \nonumber
\end{eqnarray}
In Equation (\ref{SN_ex}) $\lambda _{1}$ and $\lambda _{2}$ refer to the
eigenvalues of the differential operator in Eq. (\ref{dnx_eq}) and are given
by 
\begin{equation}
\lambda _{1,2}=\frac{1}{2L_{E}}\left( -1\pm \sqrt{1+4\frac{L_{E}^{2}}{%
\widetilde{L}_{D}^{2}}}\right) \text{;}  \label{lambda}
\end{equation}

According to standard results of GR noise theory,\cite{Vliet65} $S_{I}(0)$
is conveniently written as 
\begin{equation}
S_{I}(0)=\left( \frac{\overline{I}}{\overline{N}}\right) ^{2}\left[
S_{N}^{\infty }(0)-S_{N}^{ex}(0)\right] \text{.}  \label{SI_sol}
\end{equation}
Equation (\ref{SI_sol}), together with Eqs.(\ref{SN_inf}) and (\ref{SN_ex}),
constitutes a fully analytical and exact solution for the low frequency
current spectral density of the problem in subject and represents the main
result of the present paper. In deriving it we have not made any assumption
regarding either the condition for local space charge neutrality or the
relevance of the diffusion current. We note, that in spite of the minus sign
in Eq. (\ref{SI_sol}), $S_{I}(0)$ is a positive definite quantity, as should
be.

\begin{figure}[tbp]
\centerline{
\epsfxsize=9cm \epsffile{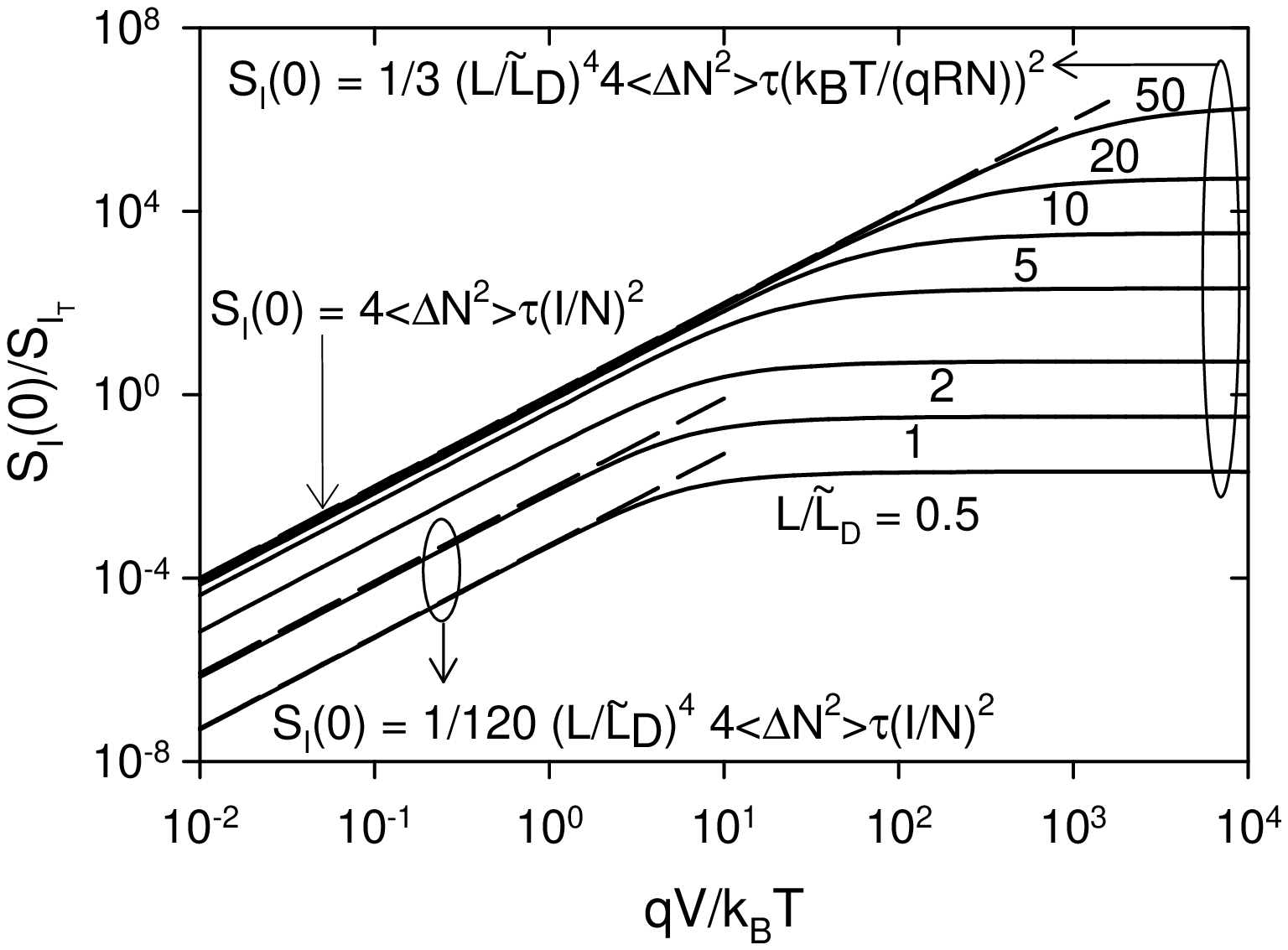}}
\caption{Spectral density of current fluctuations associated with GR noise
vs applied voltage at different sample lengths.}
\label{Fig1}
\end{figure}

To investigate the physical properties of the solution, we note that the
term $S_{N}^{ex}(0)$ vanishes for an infinitely long sample thus implying 
\begin{equation}
S_{I}^{\infty }(0)=4\left\langle \Delta N^{2}\right\rangle \tau \left( \frac{%
\overline{I}}{\overline{N}}\right) ^{2}\text{,}  \label{SI_inf}
\end{equation}
as known from simpler theories.\cite{Vliet65,Ziel86} Therefore, $%
S_{N}^{ex}(0)$ constitutes a size effect related to the finite nature of the
sample. This size effect is analyzed below by investigating the dependence
of the current spectral density upon applied voltage. The results are
summarized in Fig. \ref{Fig1} where we find convenient to normalize the
current spectral density $S_{I}(0)$ to the spectral density corresponding to
an infinite sample with thermal current $I_{T}=k_{B}T/(qR)$, i.e. $%
S_{I_{T}}= $ $4\left\langle \Delta N^{2}\right\rangle \tau \left( I_{T}/%
\overline{N}\right) ^{2}$, and the applied voltage $V$ to the thermal
voltage $V_{T}=k_{B}T/q$. With these renormalizing factors, $%
S_{I}(0)/S_{I_{T}}(0)$ as a function of $qV/(k_{B}T)$, depends only on the
dimensionless length, $L/\widetilde{L}_{D}$.

The essential result we want to stress from Fig. \ref{Fig1} is that the
spectral density saturates for high applied voltages to the value 
\begin{equation}
S_{I}^{sat}(0)=\frac{1}{3}\left( \frac{L}{\widetilde{L}_{D}}\right)
^{4}4\left\langle \Delta N^{2}\right\rangle \tau \left( \frac{k_{B}T} {qR 
\overline{N}}\right) ^{2}\text{.}  \label{SI_sat}
\end{equation}
The critical voltage for saturation $V_{sat}$ satisfies

\begin{equation}
V_{sat}=\left\{ 
\begin{array}{c}
\sqrt{40}\frac{k_{B}T}{q}\text{;\qquad for }\frac{L}{\widetilde{L}_{D}}<1 \\ 
\left( \frac{L}{\widetilde{L}_{D}}\right) ^{2}\frac{k_{B}T}{q}\qquad \text{%
for }\frac{L}{\widetilde{L}_{D}}\gtrsim 10
\end{array}
\right. \text{,}  \label{V_sat}
\end{equation}
(and an intermediate behavior for $1<L/\widetilde{L}_{D}<10$). The low bias
behavior of $S_{I}(0)$ displays also interesting features: it equals $%
S_{I}^{\infty }(0)$ for $L/\widetilde{L}_{D}>10$ and $0<V<\left( L/%
\widetilde{L}_{D}\right) ^{2}k_{B}T/q$, but it is suppressed to $1/120\left(
L/\widetilde{L}_{D}\right) ^{4}S_{I}^{\infty }(0)$ for $L/\widetilde{L}%
_{D}<1 $ and $0<V<\sqrt{40}\frac{k_{B}T}{q}$ (and an intermediate behavior
in the range $1<L/\widetilde{L}_{D}<10$).

The saturation of $S_{I}(0)$ and the dependence of the results on the ratio $%
L/\widetilde{L}_{D}$ can be explained under the assumption that both space
charge fluctuations and diffusion current play a relevant role. Indeed,
although the system is neutral in average, i.e. $\overline{n}(x)=\overline{N}%
_{t}^{+}(x)$, instantaneously space charge fluctuations can appear for which 
$\delta n(x,t)\neq \delta N_{t}^{+}(x,t)$. The relevance of these
fluctuations is determined by their interaction with the rest of charges in
the system through the long range Coulomb interaction. In this line of
reasoning for $L<\widetilde{L}_{D}$ the long range Coulomb interaction does
not play any role at all, and hence, as soon as the applied voltage is
higher than the thermal value, the low frequency current spectral density
saturates. On the other hand, when $L>\widetilde{L}_{D}$ the long range
Coulomb interaction is able to restor space charge neutrality for low
applied bias. However, as long as the applied bias is high enough as to make
the transit time $\tau _{T}=L^{2}/(\mu V)$ shorter than the renormalized
dielectric relaxation time $\widetilde{\tau }_{d}=\epsilon /[q\mu \overline{n%
}(1+\tau _{t}/\tau _{n})]$, the long range Coulomb correlations vanish and
the low frequency spectral density saturates.\cite{Gomila00} Furthermore,
the fact that the diffusion current is playing a relevant role is seen on
the fact that no shot noise is obtained in the high bias regime as it is
obtained when the diffusion current is neglected.\cite{Vliet65,Vliet75}

It is worth noting that the results presented in the present paper are
qualitatively similar to the numerical results recently reported by Bonani
and Ghione\cite{Bonani99} for a more complex GR model. This fact indicates
that saturation at high applied bias can be a general feature of GR noise.
The present work, then, offers a more complete understanding of the finite
size effects on generation-recombination noise in semiconductor devices.


Acknowledgments

Partial support from the MCyT-Spain through the Ramon y Cajal program and
project No. BFM2001-2159 and from the MADESS II is gratefully acknowledged.
Prof. F. Bonani of Torino University is acknowledged for the stimulating
discussions carried out on the subject.

\end{multicols}

\end{document}